\documentclass[%
superscriptaddress,
showpacs,
showkeys,
preprintnumbers,
nofootinbib,
 amsmath,amssymb,
aps,
]{revtex4-1}
\usepackage[english]{babel}
\usepackage{amsfonts}
\usepackage{amsmath}
\usepackage{wasysym}
\usepackage{amssymb}
\usepackage{booktabs}
\usepackage{graphicx}
\usepackage{multirow}
\usepackage[papersize={25cm,35cm}]{geometry}
\usepackage[colorlinks=true,linkcolor=blue,citecolor=green,menucolor=blue]{hyperref}
\usepackage{url}
\usepackage[latin1]{inputenc}

\def\L#1{\left\{ #1 \right\}}

\def\P#1{\left( #1 \right)}

\begin{document}
\preprint{cond-mat.mtrl-sci}
\title[Sub micron--precision sample holder for accurate re--positioning of samples in Scanning Force Microscopy]{Sub micron--precision sample holder for accurate re--positioning of samples in Scanning Force Microscopy}

\author{José Abad}
\email{jabad@um.es}
\affiliation{Dep. de F\'{i}sica Aplicada, Universidad Polit\'{e}cnica de Cartagena, E-30202, Cartagena (Spain)}
\author{Juan Francisco Gonz\'{a}lez Mart\'{i}nez}
\email{jfgm@um.es}
\author{Jaime Colchero Paetz}
\email{colchero@um.es}
\affiliation{Instituto Universitario de Investigaci\'{o}n en \'{O}ptica y
Nanof\'{\i}sica (IUOyN), Campus de Espinardo, Universidad de Murcia, E-30100
Murcia (Spain)}
\date{\today}

\begin{abstract}

Scanning Probe Microscopy allows for extreme resolution down to the atomic scale. Unfortunately, total scanning range is rather limited, therefore finding a specific position on the sample is tedious. This is an important limitation of many Scanning Probe Microscopes, in particular when the sample has to be removed for some kind of treatment and then re--allocated to characterize the same position where the previous experiment had been performed. In the present work we describe two simple and compact sub micron--precision sample holders that can be easily integrated in to a commercial Scanning Force Microscopy system. The design is based either on a traditional kinematic mounting or on self--adjustment of the sample holder and the upper piece of the piezoelectric scanner as the glue used to assemble the final setup solidifies. With these sample holders a specific sample position is automatically recovered to within about 100 nanometers, and thus well within the typical range of a piezoelectric scanner. Our experimental setup therefore allows ex--situ manipulation of the sample and SFM imaging of the same region without the aid of an optical microscope, positioning marks and tedious re--allocation.

\end{abstract}

\maketitle

\clearpage
\newpage

\section{Introduction}

Scanning Force Microscopy (SFM)~\cite{t1} is an extremely powerful tool for a variety of nanoscale research areas. Its unique ability to image, characterize different material properties as well as to modify surfaces with nanometer and even atomic resolution have attracted the interest of many researchers. However, in some cases this extreme resolution may be a disadvantage: since the surface is analyzed at very high resolution, it may not be clear if a specific spot analyzed is representative of the whole surface, or whether one has had the bad (or good) luck to explore a very special region of the surface. This problem is particularly severe when the evolution of the surface properties is to be studied as some process is applied to it ex--situ, that is, when the sample has to be removed and brought back into the SFM system. Variations of the sample observed during surface processing may then be either due to local variations of the surface properties, which is usually not what is looked for, or due to variations induced by surface processing, which is what one usually really intends to study. A particularly interesting application of precise re--allocation of the sample is Nanotomography~\cite{t2,t3}, where by successive SFM imaging and surface removal processes a nanoscale 3--dimensional characterization of the sample is obtained.

When the sample is removed for processing and then re--allocated it is usually quite difficult to find the same position, making it therefore difficult to distinguish whether a surface property is characteristic of the whole sample and/or a surface process or merely a product of a statistical variation (``noise'') of a particular region. To avoid errors induced by local variations of surface properties either the same spot has to be found after re--allocation, or a study of different regions has to be performed, which may take a considerable amount of time when a statistically significant amount of data is needed. In the case of Nanotomography, precisely the same sample position has to be recovered, otherwise the technique cannot be implemented. The same spot can be found after removal of the sample by combining an appropriate sample holder with an optical microscope combined with the SFM system, which enables targeting a specified location on the sample. The sample or the sample holder has to be marked or scratched in order to find the mark with the optical microscope and re--allocate the tip--sample position with a precision that should be significantly better than the scan range of the piezoelectric scanner used in the SFM system. After each surface processing step the mark is then found with the optical microscope, the whole SFM setup is re--aligned and successive SFM images with smaller size are taken in order to precisely re--allocate the tip--sample position~\cite{t2,t3} This method may be quite time consuming and requires the fabrication of substrates with the optical marks. Even though such substrates are also commercially available~\cite{t4,t5}, the materials of these substrates may not be the appropriate ones for a particular experiment. In order to reduce alignment time, and avoid the need for special substrates and marks, as well as the optical microscope, in the present work a sub micron--precision sample holder has been developed that is easily integrated into a commercial SFM system. Our experimental setup allows ex--situ manipulation of the sample and SFM imaging of the same region with an accurate re--positioning without the aid of an optical microscope.

\section{Experimental}

Three kinds of substrates have been used in this study: Glass cover slips, Highly Ordered Pyrolytic Graphite (HOPG) and the aluminum side of a DVD. The glass cover slips were used to prepare thin films of conducting polymers, which have been extensively studied in our group~\cite{t6}. To prepare these samples Regioregular poly--3--octylthiophene (P3OT) was purchased from Sigma--Aldrich, with 98.5\% head to tail couplings, polydispersity index $D = 2.6$ and molecular weight $M_n= 54000$. For spin--coating P3OT solutions in toluene were prepared with a concentration of 20 g/l. The second substrate, HOPG, was used to assemble small islands of sodium dodecyl sulfate (SDS). To obtain these samples, a solution of SDS was prepared in ultrapure water with a concentration of 10 $\mu$g/l.

The morphology of the samples was studied at room temperature and ambient conditions using SFM. A Nanotec Electronica SFM system with a phase locked loop (PLL)/dynamic measurement board~\cite{t7} was used with Olympus OMCL-AC-type cantilevers (nominal force constant: 2 N/m; resonance frequency: 70 kHz). Imaging was performed in non--contact dynamic SFM (NC--DSFM) using the oscillation amplitude as feedback parameter. Typical free oscillation amplitudes were 10 nm (peak to peak), and a 5 to 10\% reduction of the free oscillation amplitude was chosen as feedback parameter in order to maintain the tip--sample system in the ``attractive'' part of the interaction and avoid tip-sample contact. The PLL loop of the dynamic measurement board is enabled to keep the cantilever always at resonance and track the resonance frequency when it changes due to tip--sample interaction. As will be discussed below in more detail, for analysis of the SFM data WSxM software was used~\cite{t7,t8}.

\section{Description of the re-allocation system}

Two kinds of sample holders have been fabricated, one type is based on a traditional kinematic mounting while the other is based on self-adjustment of the sample holder and the upper piece of the piezoelectric scanner as the glue used to assemble the final setup solidifies. While the first kind of sample holder is essentially a Kelvin Coupling --and thus a kinematic mounting, see below-- the second kind of sample holder over-defines the position of the sample with respect to the piezo scanner and is thus not a kinematic mounting.

\begin{figure}[!ht]
\begin{center}
\includegraphics[width=16cm]{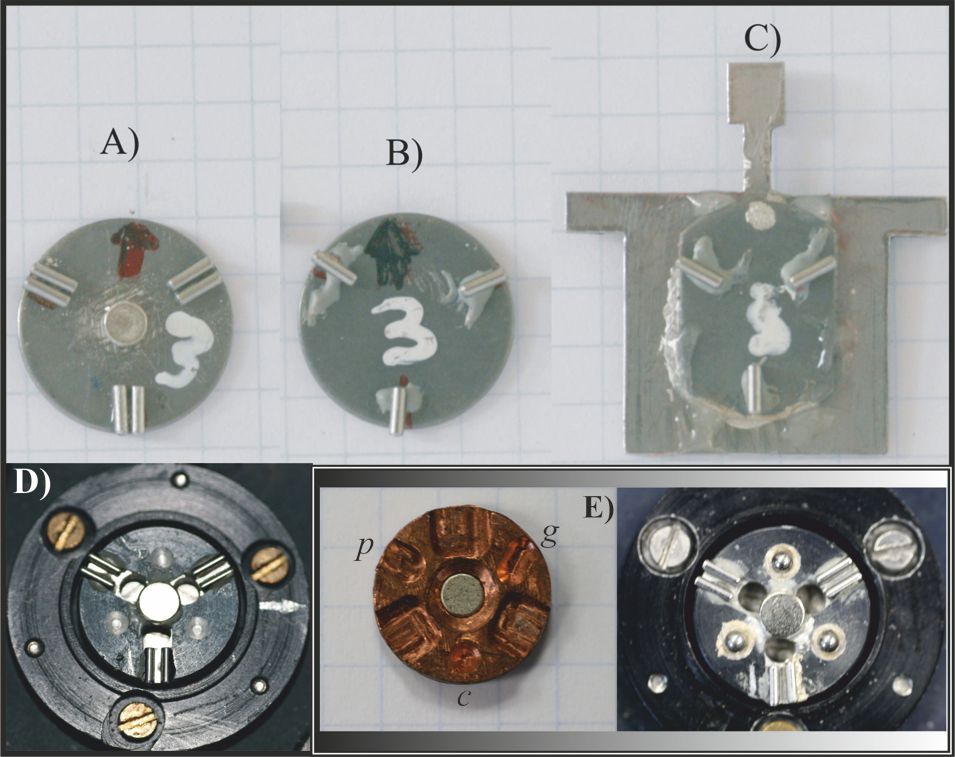}
\end{center}
\caption{ A) homemade master sample holder; B) secondary sample holder of A); C)   secondary UHV sample holder of A). D) Master holder attached to the top of the piezo scanner showing the pairs of parallel cylinders glued to the top of the piezo scanner. E) (left part) Kinematic mounting sample holder. Letters $g$, $c$ and $p$ indicate groove, cone and flat surface, respectively. E) (right part) Multifunctional  master holder.}\label{fig1}
\end{figure}

We recall that a usual kinematic mount~\cite{t9,t10,t11} is characterized by fixing all 6 degrees of freedom of one mechanical piece with respect to a second one: the tree coordinates corresponding to the centre of mass, the 2 coordinates defining the normal vector with respect to some plane, and the rotation angle around this normal vector. The most widely used kinematic mounting is the Kelvin Coupling~\cite{t12,t13}, named after Lord Kelvin. Figure \ref{fig1} E) (left part) shows our sample holder based on a Kelvin Coupling design and figure \ref{fig1} E) (right part)  the master holder, which accepts this Kelvin Coupling sample holder. Note that the master holder shown in figure \ref{fig1} E) is multifunctional; that is, in addition to the Kelving Coupling sample holder it can be also used with the second sample holder to be discussed in detail below. For positioning of the Kelvin Coupling sample holder only the tree balls on the master holder are needed. In the Kelvin coupling three balls are attached to the surface of one of the pieces which will fit into a cone (ideally a concave tetrahedron), a groove and a flat surface of the second surface. The first ball fitting into the cone ($c$ in figure \ref{fig1} E) will fix three coordinates (one relative position $\L{x,y,z}$ of pieces is fixed, but the pieces are still free to rotate), the groove ($g$ in figure \ref{fig1} E)  will fix another two (two rotation angles) and the flat surface ($p$ in figure \ref{fig1} E) will fix the last one (rotation around the axis joining the two other balls). If a mount would constrain more than the necessary six degrees of freedom the relative position of the pieces would be over--defined and perfect machining of the pieces would be necessary in order to have all mechanical boundary conditions compatible. The concept of the second sample holder is quite simple: it is based on over--defining the position of a special sample holder with respect to the SFM piezo scanner and a self--adjustment of the key elements during assembly --gluing-- of the whole setup. To be more specific, we first assemble a master holder, consisting of a 12 mm diameter disc with three pairs of parallel cylinders glued to it. The pairs of cylinders are arranged on the same radius and aligned at an angle of 120º each other, as shown in figure \ref{fig1} A) and D). The second part of our alignment system --which, were confusion may be possible, will be termed secondary sample holder-- is the sample holder itself. In order to assemble this sample holder, 3 new cylinders are positioned on the corresponding pairs of cylinders on the master piece (which has been glued previously!). Carefully, glue is applied to the disk that will be the secondary sample holder at the locations where the 3 cylinders should be positioned, and this disk is then put in contact with the 3 cylinders resting on the master piece. As the glue dries, the magnetic and/or the gravity force pushing both pieces together will align the cylinders on the secondary sample holder with respect to the pairs of cylinders on the master piece. Once the glue is dried, the 3 cylinders will be attached to the secondary sample holder in a position that should be a perfect over--defined mechanical fit. A typical sample holder fabricated in this way is shown in figure \ref{fig1} B). Note that since the relative position of the two pieces is over--defined the two pieces will only fit correctly when assembled in the precise way they were fabricated. Therefore, only one of the three positions which are obtained by 120º rotation of the sample holder will be the correct one. The two pieces are thus marked to easily find this correct alignment. Once the master piece has been fabricated, several (secondary) sample holders may be assembled that are compatible with the master piece. In particular, as shown in figure \ref{fig1} C), we have assembled a sample holder compatible with ultra high vacuum (UHV) equipment in order to combine the ex--situ UHV techniques with SFM experiments. Finally, as will be discussed in more detail below, to improve the precision of the master piece of our re-allocation system it has been integrated into the top of the piezo scanner, as shown in figure \ref{fig1} D) and E). The master holder shown in figure \ref{fig1} E) is multifunctional; that is, it can be used with the Kelvin Coupling sample holder, with the second sample holder as well as with flat conventional disks. With flat conventional disks, the position is not well defined; instead coarse $x-y$ motion is possible by inertial motion induced by the piezo scanner.

\section{Applications of the re-allocation system}

\begin{figure}[!ht]
\begin{center}
\includegraphics[width=16cm]{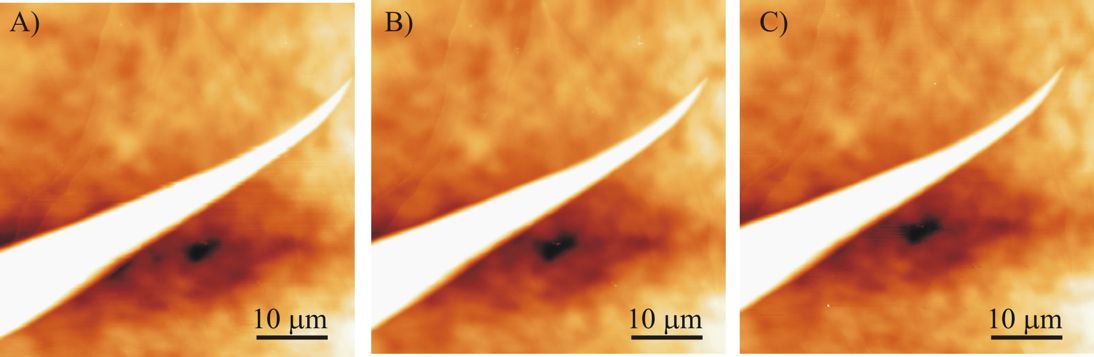}
\end{center}
\caption{ Series of SFM images of a freshly cleaved HOPG sample acquired at the same region: A) first image acquired; B) after taking out the sample and putting it back; C) after taking out the cantilever holder and putting it back. In all images  $Z=200$ nm. The sample is HOPG.}\label{fig2}
\end{figure}

To illustrate the capability of our alignment system, first a bare HOPG surface is chosen as substrate. The sample was prepared by the standard preparation method using a scotch tape to expose a freshly cleaved surface. Figure \ref{fig2} A) shows a needle like structure on the graphite surface. After the acquisition of this image the head of the microscope was lifted, the sample was taken out of the microscope setup and then put back, figure \ref{fig2} B). We find an image displacement of  $\delta = 2$ $\mu$m, where $\delta$  is the total  displacement taking into account the full displacement vector, that is, $\delta=\sqrt{\delta_x^2+\delta_y^2}$, with  $\delta_x$ and  $\delta_y$ the displacements in the $x$ (=horizontal) and $y$ (=vertical) directions. After acquisition of image \ref{fig2} B) the head of the microscope was lifted again and the cantilever holder was taken out and put back onto the head of the microscope. The image obtained after this procedure is shown in figure \ref{fig2} C), finding an image displacement $\delta = 2.5$ $\mu$m. We conclude that the SFM tip is re--positioned at almost the same region of the surface in both cases, that is, if we remove and re--position the sample, figure \ref{fig2} B), as well as if we remove and re-position the cantilever holder, figure \ref{fig2} C). The exact repositioning of the cantilever holder in the SFM head using a traditional kinematic mount is a capability of our commercial SFM head~\cite{t7}. It is therefore possible to clean the tip of the cantilever by means of UV/ozone treatment or evaporate metals or to deposit molecules on the tip, without losing the region to be studied. In the future we intend to use alignment chips~\cite{t14}, in order to allow for tip replacement without losing the precise positioning of the tip--sample system.

\begin{figure}[!ht]
\begin{center}
\includegraphics[width=16cm]{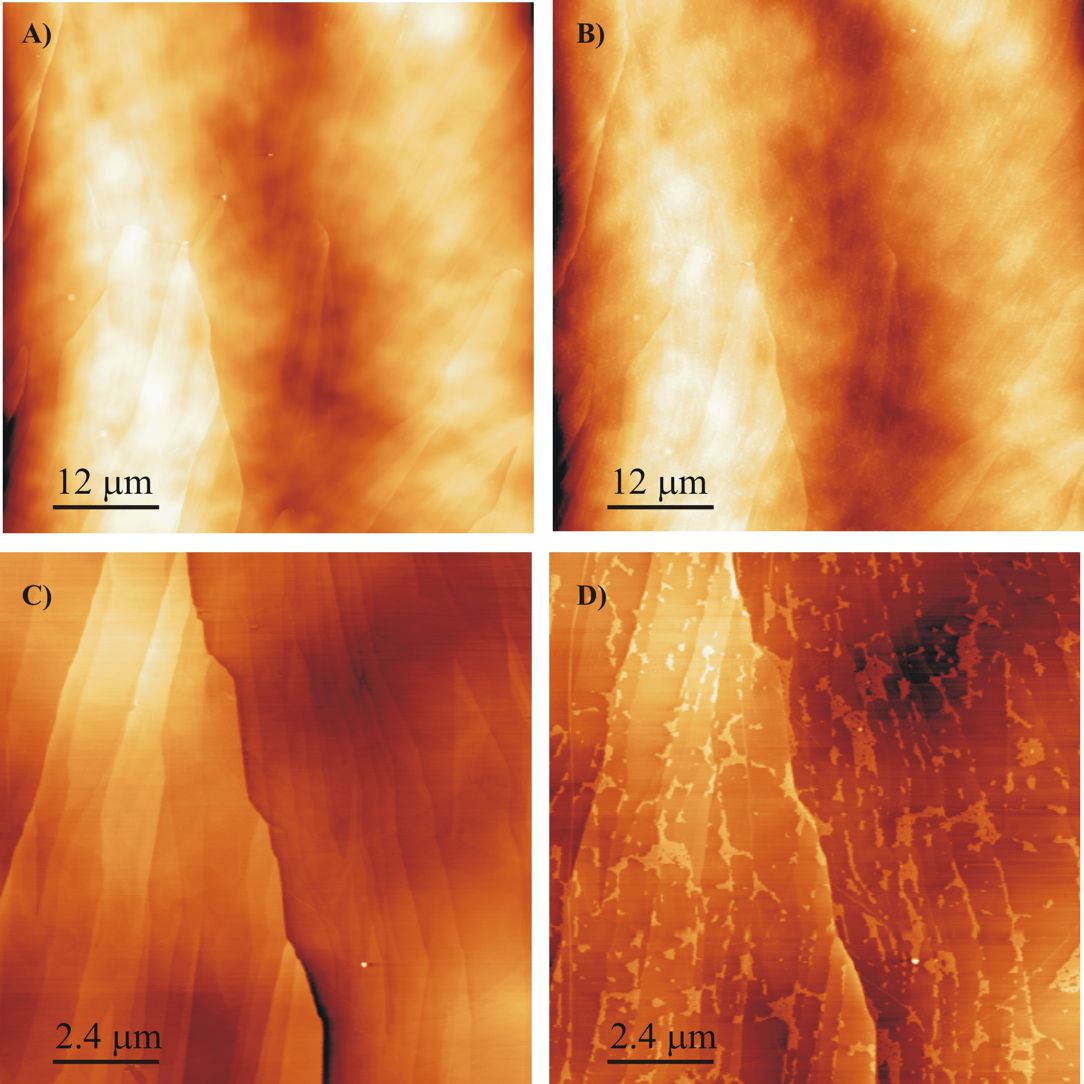}
\end{center}
\caption{SFM topographic images: A) HOPG pristine sample; B) After SDS island growth; C and D are higher resolution images of the sample, where the piezo offset voltages have been used, to precisely find the same location after re-allocation of the sample.. The scale in images A) and B) is  $Z=200$ nm, and in images C) and D) is  $Z=50$ nm.}\label{fig3}
\end{figure}

In a second experiment, the alignment was tested with a HOPG sample on which SDS islands were grown. In addition, this experiment demonstrates how the alignment system can be used to study the adsorption of materials on different substrates. The SDS islands were grown by drop casting a SDS solution prepared with ultra pure water. Figures \ref{fig3} A) and C) show the pristine HOPG surface, while figures \ref{fig3} B) and D) show the SDS islands grown on the HOPG substrate. A distance shift between images \ref{fig3} A) and 3 B) of about 3.5  $\mu$m is found. As can be seen in figures \ref{fig3} C) and D) with the additional re--positioning using the offset voltage of the high voltage applied to the piezo scanner the images can be aligned with a precision of just a few nanometers.

\begin{figure}[!ht]
\begin{center}
\includegraphics[width=16cm]{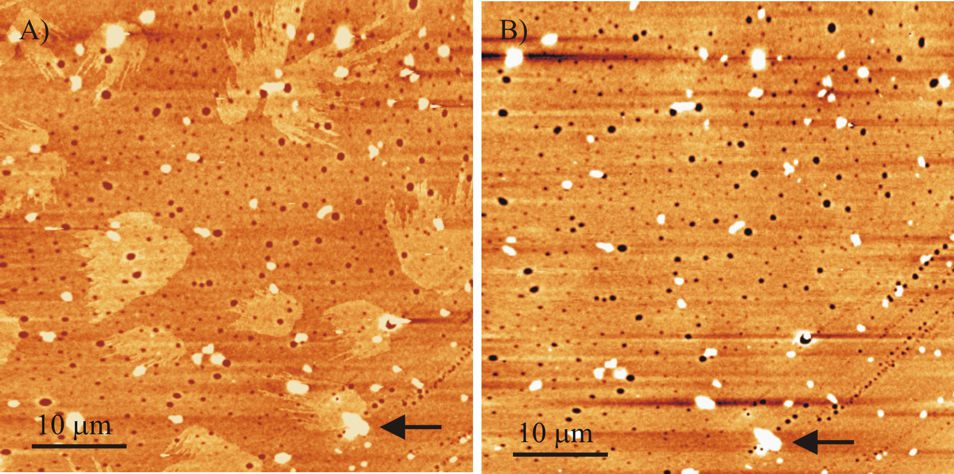}
\end{center}
\caption{SFM images of a P3OT thin film: A) pristine sample; B) after 10 minutes of Ar ion bombardment. In both images  $Z=25$ nm.}\label{fig4}
\end{figure}

The following test was carried out using P3OT polymer thin films on glass substrates glued with silver paint onto the UHV compatible sample holder. The maximum scanning size of the piezo scanner was 60 $\mu$m. Details regarding the growth and morphology of the P3OT thin films are given elsewhere~\cite{t6}. Figure \ref{fig4} A) shows the typical morphology of these films where two distinct regions are clearly distinguished: a lower region which usually covers most of the sample and a second higher region with a characteristic layered structure. The structures corresponding to the higher regions have typically one or two layers with a height of about 4.5 nm. Figure \ref{fig4} B) shows the same region after 10 minutes of Ar ion bombardment in a UHV chamber. Due to this bombardment the layered structures have been removed by the sputtering with Ar ions. A relatively large distance shift between both images of about 8  $\mu$m was found. The displacement was calculated with respect to the white feature marked in the figure with an arrow.

\begin{figure}[!ht]
\begin{center}
\includegraphics[width=16cm]{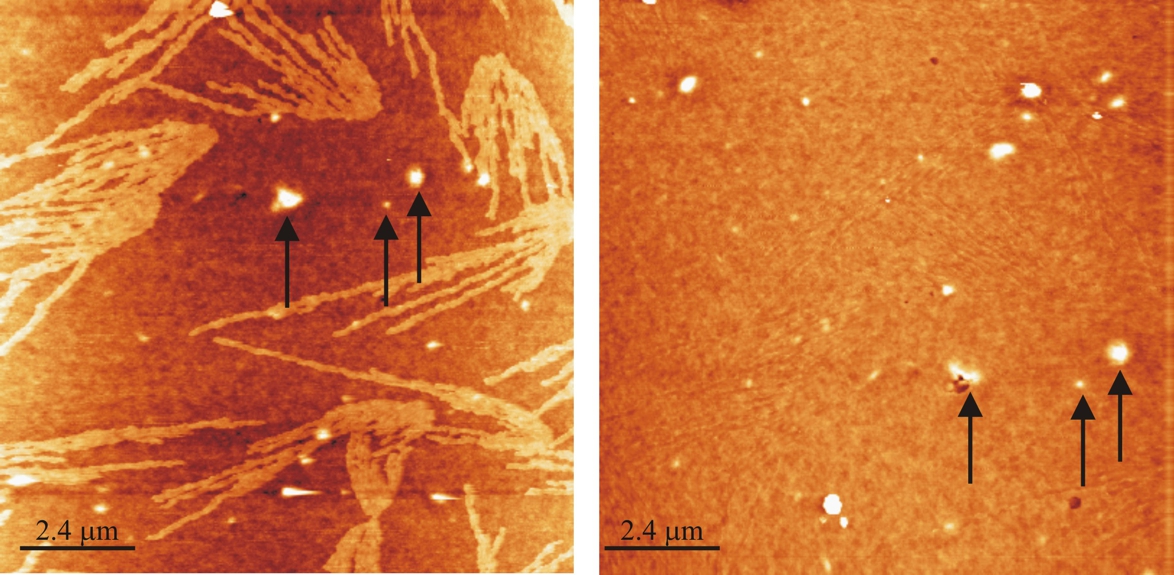}
\end{center}
\caption{SFM topographic images A) P3OT thin film pristine sample, B) After the treatment with isopropanol as discussed in the main text. The scale of both images is  $Z=20$ nm. Arrows mark the same features in both images.}\label{fig5}
\end{figure}

In a similar experiment with an equivalent P3OT thin film a drop of isopropanol was cast on a P3OT thin film prepared and imaged previously. In this experiment a piezo scanner of 12 $\mu$m scanning size, and thus more resolution, was used. Figure \ref{fig5} A) shows the pristine P3OT thin film, again the layered structures typical for these samples are clearly observed~\cite{t6}. After chemical modification of the surface with isopropanol these layered structures disappear, as observed in figure \ref{fig5} B). We note that some very faint stripes are still visible where the layered structures were before. The arrows mark the same features in both images. We find a distance shift between images \ref{fig5} A) and B) of about 4 $\mu$m.

\section{Calibration of the re-allocation system}

The examples shown so far demonstrate the working principle of the alignment system presented. In addition, these examples illustrate the variety of experiments that can be performed with such a system and the great potential of this kind of experiments. However, the precision obtained in the experiments described so far is in the few micrometers range, rather than in the submicron range as would be desirable. Two issues have limited accuracy in the experiments presented up to now. On the one hand, the master sample holder shown in figure \ref{fig1} A), is rather homemade and is fixed only magnetically onto the piezo scanner in our SFM system. Even though the magnet is attached quite firmly, a small relative movement of these two pieces when removing the sample cannot be excluded. On the other hand, all SFM images shown up to now were acquired with the SFM head on a $x-y$ mount that allows to move the tip with respect to the sample. Unfortunately, we have found that this alignment system has some small clearance that impedes precise re--alignment of the tip--sample position. To solve these two issues, on the one hand the special master sample holder shown in Figure \ref{fig1} D) was machined which is directly screwed to the top of the piezo scanner, resulting in a very solid fixation of the sample holder with respect to the piezo scanner. On the other hand, the $x-y$ mount for lateral coarse movement was substituted by a special Kelvin Coupling mounting that supports the head of the microscope with the tip and the detection system. While this kinematic mounting does not allow any more for changing the relative tip--sample position --that is, once the sample is glued to the sample holder only the same fixed location on the sample can be imaged-- it results in a much more stable and reproducible alignment of the tip with respect to the sample, as will be shown below.
With these modifications we believe that alignment error in our SFM--system is only due to:
\begin{itemize}
\item[i.-] Typical drift.
\item[ii.-] Misalignment due to coarse approach using the motorized micrometer screw of our SFM--system.
\item[iii.-] Misalignment due to removal of the microscope head.
\item[iv.-] Misalignment due to sample removal.
\end{itemize}
As discussed in more detail in the additional information, drift is negligible as compared to the other repositioning errors to be described below (350 nm  in a 34 hour movie with 480 frames; corresponding to about 0.2 nm/min).

Unfortunately the sample cannot be securely extracted from our system without some coarse withdrawal of the tip, and then removal of the head. Therefore, sample removal requires in total three different actions, each introducing some alignment error. In what follows we will analyze the positioning error of these three different actions, that is:
\begin{itemize}
\item[I.-]Only coarse withdrawal and new coarse approach.
\item[II.-]Coarse withdrawal and approach with an additional head removal and repositioning.
\item[III.-]Coarse withdrawal, head removal and sample removal then sample and head repositioning and finally coarse approach.
\end{itemize}
Assuming that the positioning errors introduced by each of the actions are statistically independent, the observed errors should add quadratically, correspondingly:
\begin{eqnarray}
&d_I^2=d_{\rm{approach--withdrawal}}^2\label{eq1}\\
&d_{II}^2=d_{\rm{head}}^2+d_I^2\label{eq2}\\
&d_{III}^2=d_I^2+d_{\rm{head}}^2+d_{\rm{sample}}^2=d_{II}^2+d_{\rm{sample}}^2\label{eq3}		
\end{eqnarray}
Therefore, we conclude that the positioning errors $d_{\rm{head}}$ and $d_{\rm{sample}}$ as determined from the previous expressions are approximately:
\begin{eqnarray}
&d_{\rm{head}}=\sqrt{d_{II}^2-d_I^2}\label{eq4}	\\						
&d_{\rm{sample}}=\sqrt{d_{III}^2-d_{II}^2}\label{eq5}					
\end{eqnarray}

\begin{figure}[!ht]
\begin{center}
\includegraphics[width=18cm]{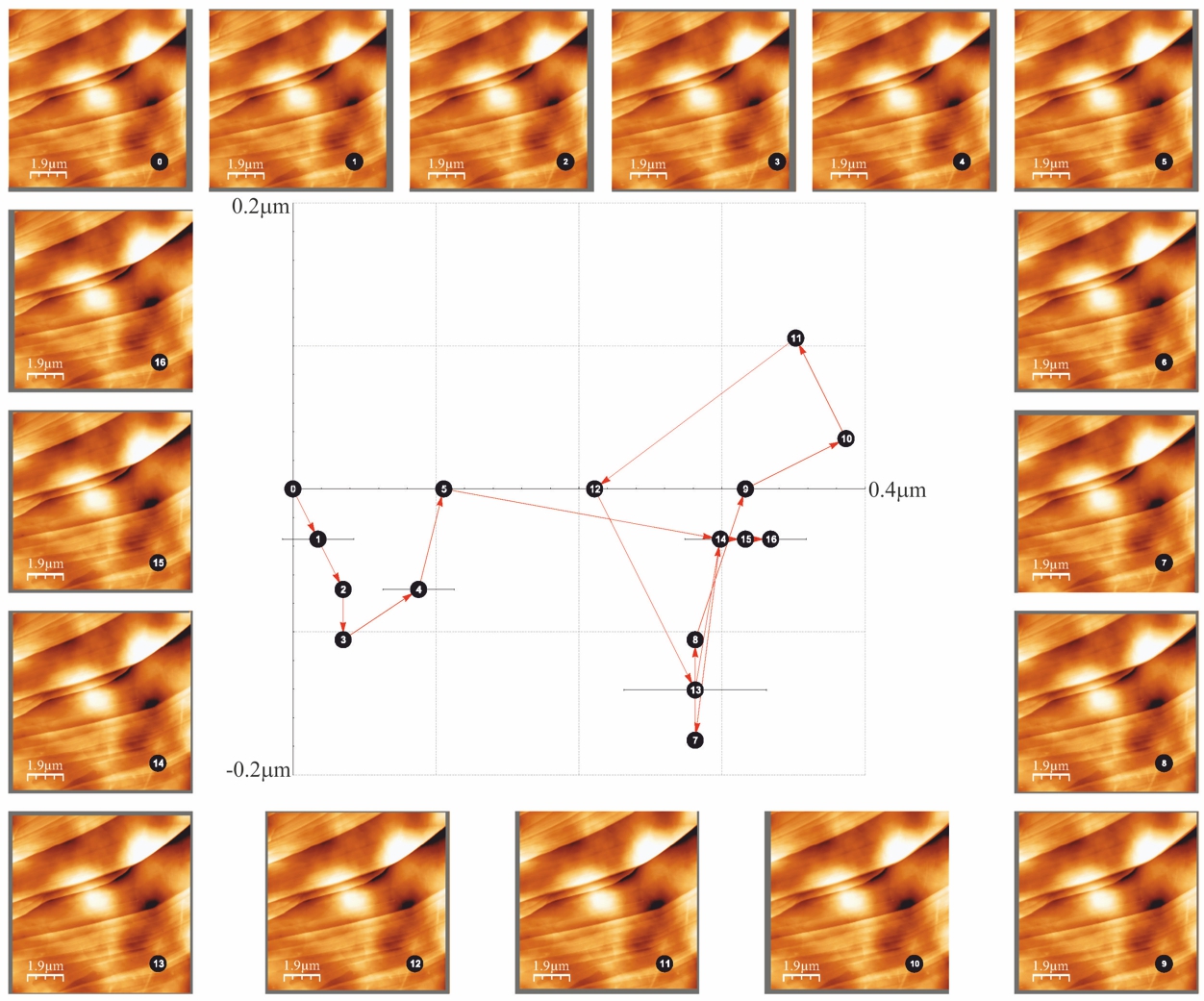}
\end{center}
\caption{SFM topography images of a cycle of sample re--positioning experiments for the non--kinematic sample holder. In total 16 individual coarse withdrawal, head and sample removal then sample and head repositioning and finally coarse approach cycles are shown. The graph in the middle of the figure shows the relative tip--sample position after each (total) re--positioning cycle. The positioning range shown corresponds to $\pm$200 nm.}\label{fig6}
\end{figure}

Another way of calculating the accuracy of the repositioning is to assume a ``random walk'' of the 2--dimensional position data (see below) and to define the accuracy of the system as the mean square deviation of the data measured from its mean position. Three different sets of experiments have been performed in order to study the alignment error corresponding to actions I, II and III. To obtain a statistically significant amount of data each removal and reallocation experiment of type I, II and III was repeated several times (typically 10 or 16). The corresponding SFM images were processed as follows: first, all images of the same data set were combined to obtain a ``movie'' with the free WSxM software~\cite{t8}. This software allows for ``movie drift correction'' by computing the cross--correlation between images and finding the position of the maximum of the cross--correlation, which directly corresponds to the offset between images of the movie. With this information a drift--corrected movie, as well as a path with the offset vectors is calculated. In the present case, the offset vectors correspond to the repositioning errors between successive removal and realignment cycles, which is the essential information to evaluate the performance of our alignment system. Figure \ref{fig6} shows all images as well as the path corresponding to the relative tip-sample position for an experiment of type III; that is: coarse tip retraction, head removal, sample removal, sample reallocation, head reallocation and coarse tip approach. The graph shows the relative tip--sample motion induced by the whole process. Table 1 of the additional information, column III$_{NK}$, lists the corresponding offset vectors calculated for each cycle. We note that for each cycle the topography and amplitude images (=feedback signal) corresponding to backward and forward scan are acquired, which allows to calculate four paths. From the relative variation of these four paths a statistical error for the determination of the tip--sample motion after each realignment step can be estimated. As discussed in the additional information, we find that when fine--tuned, the paths calculated from the error signals are more precise. For a much more detailed discussion on how these paths and their errors are calculated, see the additional information. There, the whole movies are presented, which allow a very visual analysis of the whole allocation and re--allocation process.

\begin{figure}[!ht]
\begin{center}
\includegraphics[width=18cm]{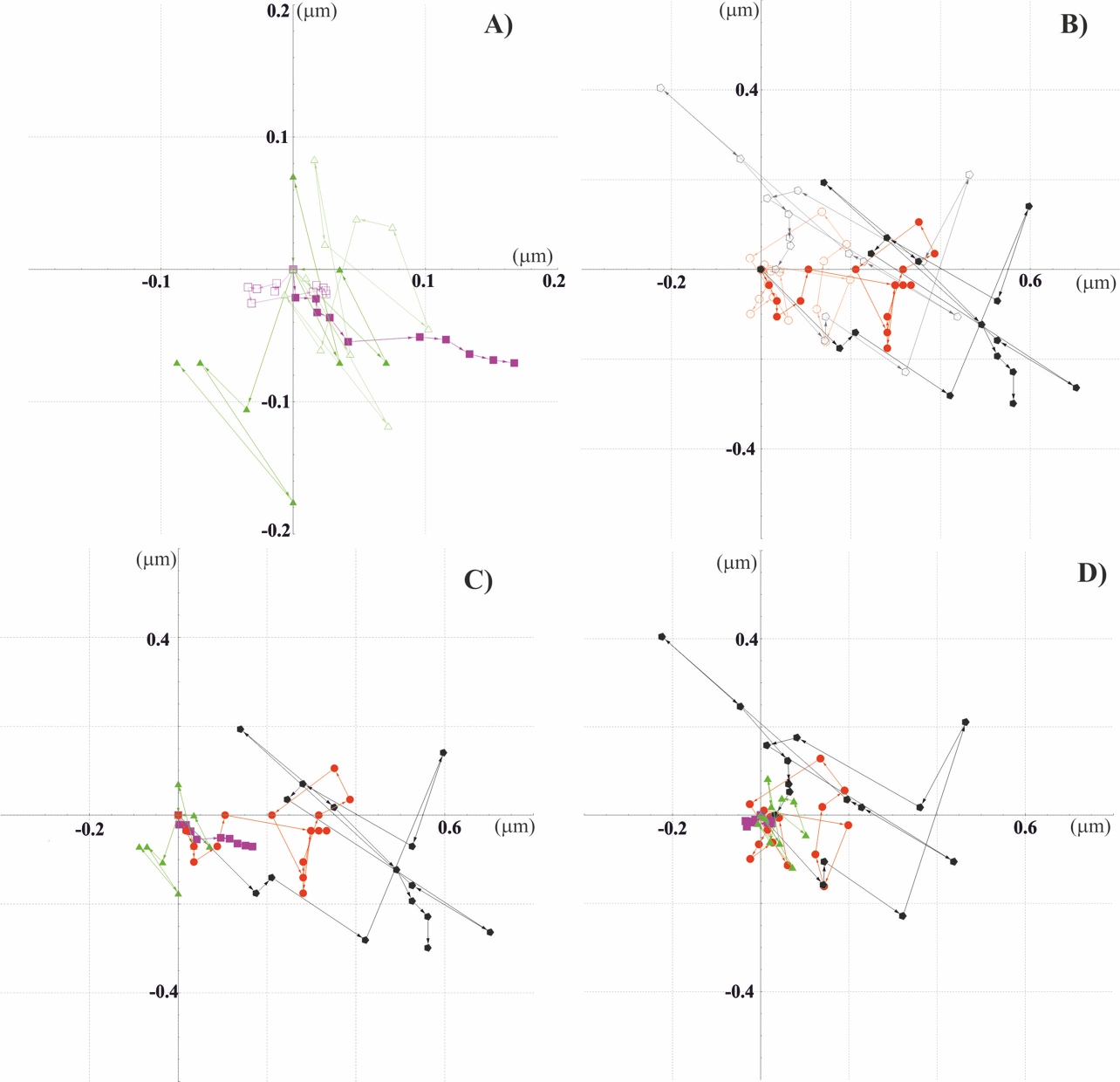}
\end{center}
\caption{Graphs summarizing the different re--positioning experiments discussed in the main text: experiments of type I (coarse approach and withdrawal),II (coarse approach and withdrawal + head removal and repositioning) and type III (sample repositioning, including coarse approach and withdrawal as well as head removal and repositioning). Graph A) shows data corresponding to experiments of type I (green color and triangle shape) and II magenta color and square shape), while graph B) shows the experiments of type III for the two sample holders, the Kelvin Coupling sample holder (black color and pentagonal shape) and the non--kinematic sample holder (red color and circular shape). In graphs A) and B), the stronger color corresponds to the measured raw data, while the corrected paths have been drawn in a lighter color and with empty shape. Graphs C) and D) show the data corresponding to all four experiments, in order to allow for an easier comparison. Graph C) corresponds to the raw data while graphs D) show the corrected paths as discussed in the main text. The color and shape coding of these graphs is the same as in graphs A) and B): green color and triangle shape for type I; magenta color and square shape for type II; black color and pentagonal shape for type III$_{KC}$  (=Kelvin Coupling sample holder) and finally red color and circular shape for type III$_{NK}$  (=non--kinematic sample holder).}\label{fig7}
\end{figure}

If the positioning error were completely uncorrelated, that is, if the precise re--allocation would be completely independent to the previous allocation -- re--allocation step, then the typical ``drunken man walk'' would be expected for the path followed by the relative tip--sample position in any of the experiments of type I, II and III. From the data acquired in some situations we find that this is not generally the case since the data has a clear directionality; in particular for experiments of type I --only coarse withdrawal and new coarse approach-- (see, for example, figure \ref{fig7} A). For this latter case of experiments we attribute this directionality to a movement induced by the screw that moves the SFM head for tip--sample approach and retraction. Indeed, when the screw moves, it also rotates, which induces a torque at the contact point between the head and the screw. This torque will tend to move the head from its ideal position in a non--statistical way, and this could be the cause for the directionality observed. We have not found a direct cause for the directionality in the other type of experiments (II and III), however, since experiments II and III need also tip--sample approach and withdrawal (type I), the directionality of experiments of type I will always also affect these other two cases. In order to subtract this directionality from the data, in addition to the raw data we compute corrected paths were this directionality is removed. Technically, this is done by fitting lines $l_x[i]$ and $l_y[i]$ to the data $(i, x[i])$ and $(i, y[i])$, where $i$ is the number of the step and $x[i]$ and $y[i]$ are the horizontal and vertical position measured after each step. The corrected paths $p_c=\L{\P{x_c[1],y_c[1]},\ldots,\P{x_c[n],y_c[n]}}$ are then computed by calculating the distance from the lines to the corresponding data: $x_c[i]=l_x[i]- x[i]$ and $y_c[i]=l_y[i]- y[i]$, where $l_x[i]$ and $l_y[i]$ are the lines interpolating the horizontal and vertical displacements. As will be seen below, these corrected paths show a much more ``drunken man walk'' behavior as compared to the original data with the raw displacement curves.

In order to compare the performance of the two sample holders, the experiment of type III --coarse withdrawal, head and sample removal then sample and head repositioning and finally coarse approach-- was performed twice, once for each sample holder. The experiments of type I --coarse withdrawal and new coarse approach-- and of type II --coarse withdrawal and approach with an additional head removal and repositioning-- were only repeated once, since no action with the sample is involved, therefore the statistical properties of the corresponding paths should not vary if a different sample holder is used, The corresponding images (and movies) and raw data are presented in the additional information; and the processed curves are shown in figure \ref{fig7}. The upper graphs in this figure shows raw data corresponding to experiments of type I and II (left, figure \ref{fig7} A) and of experiments of type III with the two different sample holders (right, figure \ref{fig7} B). The lower graphs show the data corresponding to all four experiments, the left graph corresponds to the raw paths while the right graphs show the corrected curves as discussed above. We clearly observe that the positioning errors add, that is, each additional action (coarse tip approach and withdrawal, head removal and re-positioning and sample removal and re-positioning) leads to larger steps in the corresponding paths. This implies that each action leads to inaccuracies that are not negligible in the total behavior. To be more specific, the following mean step size and mean position and mean step size are calculated for the different type of experiments:
\begin{itemize}
\item[I]Only coarse withdrawal and new coarse approach:

mean step size $s_I = 13\pm13$ nm, mean position $p_I = (80,50)$ nm, mean square deviation of position $\sigma_I =(  x , y )=(60,20)$ nm, with modulus $|\sigma_{I}|\approx 65$ nm.
\item[II]Coarse withdrawal and approach with an additional head removal and repositioning:

mean step size $s_{II} = 90\pm40$ nm, mean position $p_{II} = (10,-50)$ nm, mean square deviation of position  $\sigma_{II}=(  x , y )=(40,70)$ nm, with modulus $|\sigma_{II}|\approx 80$ nm.
From this data we estimate, according to relation (4), the following values for the accuracy of the re-positioning of the head:

mean step size $s_{\rm{Head}} = 90\pm40$ nm, mean square deviation $|\sigma_{\rm{Head}}|\approx50$ nm.
\item[III$_{KC}$]	For the case of the Kelvin Coupling sample holder: Coarse withdrawal, head and sample removal then sample and head repositioning and finally coarse approach:

mean step size $s_{KC} = 230\pm170$ nm, mean position $p_{KC} = (440,-70)$ nm, mean square deviation of position  $\sigma_{KC} = (  x , y )=(180,160)$ nm, with modulus $|\sigma_{KC} |\approx$ 240 nm.
From this data we estimate, according to relation (5), the following values for the accuracy of the re-positioning of this sample holder:

mean step size $s_{KC0} = 210\pm200$ nm, mean square deviation  $\sigma_{KC0}\approx225$ nm.
\item[III$_{NK}$]	For the case of the non--kinematic (NK) sample holder: Coarse withdrawal, head and sample removal then sample and head repositioning and finally coarse approach:

mean step size $s_{NK} = 90\pm50$ nm, mean position $p_{NK} = (200,-60)$ nm, mean square deviation of position  $s_{NK} = (  x , y )=(120,60)$ nm, with modulus $|\sigma_{NK}|\approx125$ nm.
From this data we estimate, according to relation (5), the following values for the accuracy of the re-positioning of this sample holder:

mean step size $s_{NK0} = 0\pm60$ nm, mean square deviation  $\sigma_{NK0}\approx95$ nm.
\end{itemize}

From the results obtained, we conclude that most actions, and in particular the coarse approach and retraction of the tip, a clear directionality is observed, which renders the paths ``non-random walks''. If this directionality is subtracted as discussed above, the paths shown in figure \ref{fig7} D) have a much more ``random walk'' behavior as compared to the raw curves obtained. For the two different sample holders we obtain a mean positioning error (mean square positioning accuracy) of $s_{KC} = 225$ nm in case of the Kelvin Coupling and $s_{NK} = 95$ nm in case of the non--kinematic mounting taking into account the total re--positioning process (=coarse withdrawal, head and sample removal then sample and head repositioning and finally coarse approach) and thus well within the sub--micron regime. Interestingly, we find that the specific non--kinematic mounting described in this work has an almost twofold better (total) accuracy ($|\sigma_{NK}|\approx 125$ nm vs. $|\sigma_{KC}| = 240$ nm) as compared to a traditional Kelvin Coupling. For the precision of the single sample re--positioning we estimate a value of  $\sigma_{KC0}\approx225$ nm for the traditional Kelvin Coupling and  $\sigma_{NK0}\approx95$ nm.

\section{Conclusions}

A SFM sample holder for precise, quick and reproducible positioning of the sample with respect to the tip has been developed. This setup therefore allows investigation of a sample, removal of the sample from the SFM--setup for ex-situ modification and subsequent precise re-allocation without the aid of optical microscopy and positioning marks. Different prototypes of this system have been fabricated and tested. In particular, two different type of sample holders have been analyzed in detail, one based on a classical kinematic mount design (Kelvin Coupling) and another one based on a non--kinematic mount. Interestingly, we find that the latter mount has a better accuracy. Nevertheless a final re--positioning accuracy in the hundred nanometer range is obtained for both systems. We are convinced that, as shown with some representative examples in the present work, this sample holder design allows for a variety of interesting experiments, ranging from Nanotomography to a variety of studies where the evolution of nanoscale properties is characterized as some ex-situ process is applied to the sample.

\section{Acknowledgments}

The authors thank Nanotec Electronica for providing the top of the piezo scanner as well as the head of the microscope with the Kelvin coupling. In particular, the authors acknowledge L. Colchero for his comments and suggestions and I. Horcas for this help with the ``drift correction'' algorithm of the WSXM software. We also acknowledge the work and patience of Juan Francisco Miñarro from the workshop of the Universidad de Murcia, who fabricated the two ``professional'' sample holders. We also appreciate the comments and suggestions of the referee to our first version on the manuscript. This research has been financed by the ``Comunidad Autónoma de la Región de Murcia'' through the project ``Células solares orgánicas: de la estructura molecular y nanométrica a dispositivos operativos macroscópicos'' as well as by the Ministerio de Ciencia e Innovación (MICINN, Spain) through the projects ``Force For Future'' (CONSOLIDER program, CSD2010-00024) and ``Propiedades Nanométricas de Células Solares Orgánicas'' (MAT2010-21267-C02-01). In addition, JFGM acknowledges his scholarships from the Ministerio de Ciencia e Innovación (FPU-programm). This work has also been co-funded by the European Union.

\clearpage
\newpage

\section{Figures and captions}

\end{document}